\documentclass[lettersize,journal]{IEEEtran}

\usepackage{float}
\usepackage{color}
\usepackage{graphicx}
\usepackage{mathtools}
\usepackage{amsmath,amssymb,amsthm}

\usepackage[caption=false,subrefformat=parens,labelformat=parens]{subfig}
\usepackage{balance}
\usepackage[noadjust]{cite}
\usepackage{epstopdf}
\usepackage{psfrag,pst-pdf}
\usepackage[nolist]{acronym}

\usepackage{algcompatible}
\usepackage{algorithm}
\usepackage{dsfont}
\usepackage{textgreek}
\usepackage{tablefootnote}
\usepackage{threeparttable}

\usepackage{algpseudocode}

\makeatletter
\newcounter{phase}[algorithm]
\newlength{\phaserulewidth}

\makeatother

\usepackage{hyperref}

\acrodef{2D}{two-dimensional}
\acrodef{3D}{three-dimensional}
\acrodef{4G}{4th generation}
\acrodef{5G}{5th generation}
\acrodef{6G}{sixth generation}
\acrodef{ADT}{angle diversity transmitter}
\acrodef{ADR}{angle diversity receiver}
\acrodef{ANSI}{American national standards institute }
\acrodef{AP}{access point}
\acrodef{AR}{augmented reality}
\acrodef{AWGN}{additive white Gaussian noise}
\acrodef{AWGR}{arrayed waveguide grating router}
\acrodef{BER}{bit error ratio}
\acrodef{CDF}{cumulative distribution function}
\acrodef{CSI}{channel state information}
\acrodef{CPC}{compound parabolic concentrator}
\acrodef{CoMB}{coordinated multi-beam}
\acrodef{CoMB{-}JT}{coordinated multi-beam joint transmission}
\acrodef{CoMP}{coordinated multi-point}
\acrodef{DC}{direct current}
\acrodef{DCO{-}OFDM}{DC-biased optical OFDM}
\acrodef{DFB}{distributed feedback laser}
\acrodef{DMT}{discrete multitone}
\acrodef{EE}{energy efficiency}
\acrodef{EGC}{equal gain combining}
\acrodef{FEC}{forward error correction}
\acrodef{FFT}{fast Fourier transform}
\acrodef{FOV}{field of view}
\acrodef{FSO}{free space optical}
\acrodef{IBI}{inter-beam interference}
\acrodef{IEC}{International electrotechnical commission}
\acrodef{IFFT}{inverse fast Fourier transform}
\acrodef{IM{/}DD}{intensity modulation and direct detection}
\acrodef{IM{-}DD}{intensity modulation and direct detection}
\acrodef{IR}{infrared}
\acrodef{ISI}{inter-symbol interference}
\acrodef{ICI}{inter-cluster interference}
\acrodef{JT}{joint transmission}
\acrodef{LD}{laser diode}
\acrodef{LED}{light-emitting diode}
\acrodef{LiFi}{light fidelity}
\acrodef{MIMO}{multiple input multiple output}
\acrodef{MPE}{maximum permissible exposure}
\acrodef{MPTP}{maximum permissible transmit power}
\acrodef{MHP}{most hazardous position}
\acrodef{MRC}{maximum ratio combining}
\acrodef{MUI}{multi-user interference}
\acrodef{NIR}{near infrared}
\acrodef{NOMA}{non-orthogonal multiple access}
\acrodef{OOK}{on-off keying}
\acrodef{PAM}{pulse amplitude modulation}
\acrodef{PSD}{power spectral density}
\acrodef{PD}{photodiode}
\acrodef{OFDM}{orthogonal frequency division multiplexing}
\acrodef{OFDMA}{orthogonal frequency division multiple access}
\acrodef{OCC}{optical camera communication}
\acrodef{OW}{optical wireless}
\acrodef{OWC}{optical wireless communication}
\acrodef{QAM}{quadrature amplitude modulation}
\acrodef{RF}{radio frequency}
\acrodef{RIN}{relative intensity noise}
\acrodef{RSS}{received signal strength}
\acrodef{SDMA}{space division multiple access}
\acrodef{SIC}{successive interference cancellation}
\acrodef{SiPM}{silicon photomultiplier}
\acrodef{SMF}{single mode fiber}
\acrodef{SINR}{signal-to-interference-plus-noise ratio}
\acrodef{SNR}{signal-to-noise ratio}
\acrodef{Tb{/}s}{Terabit/s}
\acrodef{TIA}{transimpedance amplifier }
\acrodef{TEM}{transverse electromagnetic}
\acrodef{UHD}{ultra-high-definition}
\acrodef{VCSEL}{vertical cavity surface emitting laser}
\acrodef{VLC}{visible light communication}
\acrodef{VR}{virtual reality}
\acrodef{WiFi}{wireless fidelity}

\acrodef{ADC}{analog-to-digital conversion}
\acrodef{DAC}{digital-to-analog conversion}
\acrodef{CPU}{central processing unit}
\acrodef{BS}{base station}
\acrodef{CP}{circuit power}
\acrodef{ETP}{effective transmit power}
\acrodef{LIV}{light-current-voltage}
\acrodef{PA}{power amplifier}
\acrodef{PC}{power consumption}
\acrodef{UE}{user equipment}

\acrodef{GoB}{grid-of-beam}
\acrodef{SBC}{static beam clustering}
\acrodef{DBC}{dynamic beam clustering}
\acrodef{CMOS}{complementary metal-oxide semiconductor}
\acrodef{DSP}{digital signal processing}
\acrodef{GOPS}{giga operations per second}
\acrodef{GSPS}{giga samples per second}
\acrodef{FinFET}{fin field-effect transistor}
\acrodef{FOM}{figure of merit}
\acrodef{SNDR}{signal-to-noise-and-distortion ratio}
\acrodef{ENOB}{effective number of bits}
\acrodef{AC}{alternating current}
\acrodef{SRAM}{static random access memory}
\acrodef{FPGA}{field-programmable gate array}
\acrodef{CPE}{computational power efficiency}
\acrodef{QoS}{quality of service}
\acrodef{BFS}{breadth-first search}
\acrodef{BDMA}{beam division multiple access}
\acrodef{AI}{artificial intelligence}
\acrodef{IoT}{Internet of Things}
\acrodef{TV}{television}

\acrodef{SDI}{serial digital interface}
\acrodef{NRZ}{non-return-to-zero}
\acrodef{NRZI}{non-return-to-zero inverted}
\acrodef{OOK}{on-off keying}
\acrodef{NRZ{-}OOK}{non-return-to-zero on-off keying}
\acrodef{BPSK}{binary phase-shift keying}
\acrodef{HD}{high-definition}
\acrodef{FHD}{full HD}

\acrodef{MM}{multi-mode}
\acrodef{SM}{single-mode}
\acrodef{CP}{cyclic prefix}
\acrodef{CLT}{central limit theorem}
\acrodef{FF}{far field}
\acrodef{FWHM}{full-width at half-maximum}
\acrodef{AWG}{arbitrary waveform generator}
\acrodef{SPS}{samples per symbol}
\acrodef{BT}{bias tee}
\acrodef{IL}{insertion loss}
\acrodef{ACL}{aspheric condenser lens}
\acrodef{NA}{numerical aperture}
\acrodef{CA}{clear aperture}
\acrodef{ARC}{anti-reflective coating}
\acrodef{FC{/}PC}{Ferrule Connector/Physical Contact}
\acrodef{VGA}{variable gain amplifier}
\acrodef{BLER}{block error rate}
\acrodef{LTE}{long term evolution}
\acrodef{HH}{Hughes-Hartogs}
\acrodef{MCS}{modulation and coding scheme}
\acrodef{HARQ}{hybrid automatic repeat request}
\acrodef{SMD}{surface‑mount device}
\acrodef{GaN}{gallium nitride}
\acrodef{WDM}{wavelength division multiplexing}
\acrodef{PIN}{positive-intrinsic-negative}
\acrodef{InGaAs}{indium gallium arsenide}
\acrodef{VIS}{Vertically Integrated Systems}
\acrodef{SMPTE}{Society of Motion Picture and Television Engineers}
\acrodef{MA}{multi-aperture}
\acrodef{UI}{unit interval}
\acrodef{BNC}{bayonet Neill–Concelman}
\acrodef{BPI}{broadband pulse inverter}
\acrodef{CEQ}{cable equalizer}
\acrodef{HDMI}{high-definition multimedia interface}
\acrodef{PRBS}{pseudorandom binary sequence}
\acrodef{BPSK}{binary phase-shift keying}
\acrodef{WSS}{wide-sense stationary}
\acrodef{PNA}{performance network analyzer}
\acrodef{PLL}{phase-locked loop}
\acrodef{PAL}{phase alternating line}
\acrodef{OSC}{oscilloscope}
\acrodef{KPI}{key performance indicator}
\acrodef{URLLC}{ultra-reliable and low-latency communications}
\acrodef{XR}{extended reality}
\acrodef{MR}{mixed reality}
\acrodef{BBC}{British broadcasting corporation}
\acrodef{DASH}{dynamic adaptive streaming over HTTP}
\acrodef{HTTP}{hypertext transfer protocol}
\acrodef{IP}{Internet protocol}
\acrodef{HDMI}{high-definition multimedia interface}
\acrodef{SD}{standard definition}
\acrodef{QoE}{quality of experience}
\acrodef{CS}{compressed sensing}
\acrodef{SFP}{small form-factor pluggable}
\acrodef{SDR}{software-defined radio}
\acrodef{MZM}{Mach–Zehnder modulator}
\acrodef{PAT}{pointing, acquisition, and tracking}
\acrodef{UWOC}{underwater wireless optical communication}
\acrodef{APD}{avalanche photodiode}
\acrodef{PSK}{phase shift keying}
\acrodef{VGA}{video graphics array}
\acrodef{LOS}{line-of-sight}
\acrodef{DP}{DisplayPort}
\acrodef{IP}{Internet protocol}
\acrodef{NDI}{network device interface}
\acrodef{IT}{information technology}

\begin{document}

\title{Real-Time Transmission of Uncompressed High-Definition Video Via A VCSEL-Based Optical Wireless Link With Ultra-Low Latency}


\author{\IEEEauthorblockN{Hossein~Kazemi, Isaac~N.~O.~Osahon, Tiankuo~Jiao, David~Butler, Nikolay~Ledentsov Jr., Ilya~Titkov, Nikolay~Ledentsov, and Harald~Haas}
\thanks{
	\indent Hossein Kazemi \textit{(corresponding author)}, Isaac N. O. Osahon, Tiankuo~Jiao, and Harald Haas are with the LiFi Research and Development Center (LRDC), Electrical Engineering Division, Department of Engineering, University of Cambridge, Cambridge CB3 0FA, United Kingdom. \\
    \indent David Butler is with the BBC Research \& Development, London W12 7TQ, United Kingdom. \\
    \indent Nikolay Ledentsov Jr., Ilya Titkov, and Nikolay Ledentsov are with VI Systems GmbH (VIS), Berlin 10623, Germany.}}
\maketitle

\begin{abstract}
Real-time transmission of high-resolution video signals in an uncompressed and unencrypted format requires an ultra-reliable and low-latency communications (URLLC) medium with high bandwidth to maintain the quality of experience (QoE) for users. We put forward the design and experimental demonstration of a high-performance laser-based optical wireless communication (OWC) system that enables high-definition (HD) video transmission with submillisecond latencies. The serial digital interface (SDI) output of a camera is used to transmit the live video stream over an optical wireless link by directly modulating the SDI signal on the intensity of a $940$~nm vertical cavity surface emitting laser (VCSEL). The proposed SDI over light fidelity (LiFi) system corroborates error-free transmission of full HD (FHD) and 4K ultra-high-definition (UHD) resolutions at data rates of $2.97$~Gb/s and $5.94$~Gb/s, respectively, with a measured end-to-end latency of under $35$~ns. Since SDI standards support various video formats and VCSELs are high-bandwidth and low-power devices, this presents a scalable and inexpensive solution for wireless connectivity between professional broadcast equipment using off-the-shelf SDI components.
\end{abstract}


\section{Introduction} \label{Sec:Introduction}
Real-time and low-latency communications are among the key performance indicators for \ac{6G} networks to enable mission-critical applications, including autonomous systems, remote surgery, and industrial automation \cite{Hazra2024URLLC6G}. Hence, \ac{6G} aims to significantly advance \ac{URLLC} to meet stringent requirements for reliability and responsiveness, targeting near-zero packet loss and submillisecond latencies \cite{CWWang2023OnTheRoadTo6G}. Emerging applications such as multi-sensory \ac{XR} technologies, including \ac{AR}, \ac{VR}, and \ac{MR}, primarily rely on \ac{URLLC} to ensure seamless real-time user experiences through immersive high-fidelity interactions without causing motion sickness, lag, or disorientation \cite{CWWang2023OnTheRoadTo6G}.

A key application of \ac{URLLC} is live \ac{TV} broadcasting, where minimal latency and high reliability are crucial to guarantee real-time content delivery and uninterrupted viewer experience. \ac{BBC} has recently launched a low-latency streaming trial on \ac{BBC} iPlayer, aiming to reduce the delay between live broadcasts and online streaming. The trial evaluates low-latency dynamic adaptive streaming over the \ac{HTTP} with the objective of bringing streaming delays closer to broadcast levels while maintaining high \ac{QoE} for viewers \cite{Poole2025LowLatencyStreaming}. There are various digital video interfaces for the reliable transmission of high-quality multimedia content over cable connections, including \ac{HDMI}, \ac{DP}, \ac{IP}-based interfaces such as \ac{NDI}, and \ac{SDI} \cite{Watkinson2004DigitalInterface}. While \ac{HDMI} and \ac{DP} dominate consumer electronics and desktop video delivery applications, \ac{NDI}, which runs over Gigabit Ethernet, is mainly adopted in \ac{IT} systems \cite{Watkinson2004DigitalInterface}. Among these interfaces, \ac{SDI} remains the preferred choice in traditional broadcast infrastructures due to its robustness, low latency, and ability to support long distance transmission over coaxial or fiber-optic cables \cite{Watkinson2004DigitalInterface}. In fact, \ac{SDI} is a family of digital video interfaces that was first standardized in 1989 by the \ac{SMPTE} to transmit uncompressed, unencrypted digital video signals in broadcast \ac{TV} studios and post-production facilities \cite{Watkinson2004DigitalInterface}. Existing \ac{SMPTE} \ac{SDI} standards support a wide range of video formats and bit rates from \ac{SD} video at $270$~Mb/s to 8K \ac{UHD} video at $23.76$~Gb/s \cite{SMPTE259,SMPTE424,SMPTE2081,SMPTE2082,SMPTE2083}. \ac{SDI} ports are commonly available in professional video equipment such as cameras, recorders, vision mixers, desktop computers, monitors, among others. Despite widespread adoption of \ac{SDI}, it offers limited flexibility to system designers, as cables need to be deployed between the equipment.

Alternatively, real-time wireless video links can be realized using \ac{OWC} \cite{Wei2021,JLi2023Experimental,Jeon2023FSO6G,AlHalafi2017,MKong2022RealTime}. In \cite{Wei2021}, Wei \textit{et al.} implemented a real-time, multi-user optical wireless video transmission system using a micro‑\ac{LED} bulb with two $50$~{\textmu}m \ac{GaN} {\textmu}\ac{LED} chips driven by a \ac{FPGA} board based on \ac{QAM}-\ac{OFDM}. At the receiver side, each user is equipped with a \ac{PD} and \ac{FPGA}-based electronics to demodulate the \ac{QAM}-\ac{OFDM} signals. The setup achieved a combined data rate of $105.54$~Mb/s over a $2$~m free-space link with visual video fidelity and a measured end-to-end latency of $2$~s. In \cite{JLi2023Experimental}, Li \textit{et al.} experimentally demonstrated a \ac{FSO} video transmission system using a laser source operating at $1550$~nm and \ac{CS}. The system transmits compressed video frames at $20$~frame/s via a simulated indoor atmospheric turbulence channel over a link distance of $2$~m, achieving data rates of up to $3.125$~Gb/s with a minimum \ac{BER} of $1.079\times10^{-11}$, using an offline image reconstruction algorithm. In \cite{Jeon2023FSO6G}, Jeon \textit{et al.} developed a real-time \ac{FSO} system prototype to assess the feasibility of high-resolution video transmission over long distances for \ac{6G} networks. Their system integrates an \ac{FPGA}-based \ac{SDR} platform with a \ac{MZM}-based channel emulator to mimic a $20$~km terrestrial \ac{FSO} link under realistic atmospheric conditions. The authors demonstrated real-time 4K \ac{UHD} video transmission (i.e., $3840\times2160$ pixels transmitted at $60$~frame/s) with a data rate of $35$~Mb/s using \ac{PAM}-4 modulation. By employing spatial selective filtering and a sampling-based \ac{PAT} technique to enhance signal quality, their prototype achieved a \ac{BER} of about $10^{-4}$ while maintaining reliable performance even under varying channel conditions. In \cite{AlHalafi2017}, Al-Halafi \textit{et al.} presented an \ac{UWOC} system based on a directly modulated $520$~nm \ac{LD} at the transmitter and an \ac{APD} at the receiver, supporting real-time video transmission over underwater channels up to $5$~m. The system used \ac{QAM} and \ac{PSK} modulation schemes, achieving data rates up to $26.8$~Mb/s with a \ac{BER} as low as $1.2\times10^{-9}$, while transmitting a video resolution of $640\times480$ pixels at $30$~frame/s with an end-to-end latency under $1$~s. Furthermore, in \cite{MKong2022RealTime}, Kong \textit{et al.} proposed an underwater surveillance system, supporting real-time streaming of $1920\times1080$ \ac{FHD} video at $30$~frame/s. The underlying \ac{UWOC} system employed a pair of high-power wide-beam \ac{LED} transmitters and wide \ac{FOV} \ac{APD} receivers without optical collimation or focusing elements in order to establish robust \ac{LOS} links in dynamic underwater environments. The system achieved stable performance over $5$~m in an outdoor pool with an end-to-end latency of about $250$~ms, making it suitable for real-time underwater monitoring applications.

In this paper, we put forward a novel \ac{SDI} over \ac{LiFi} system for real-time wireless video transmission with a submillisecond latency. We design a high-performance laser-based \ac{OWC} system to implement direct modulation of the \ac{SDI} output of a camera over the optical intensity of a \ac{VCSEL}. The proposed system utilizes off-the-shelf \ac{SDI} devices for real-time signal acquisition and conditioning, offering a low-cost and scalable solution for wireless connectivity between professional video equipment. To the best of our knowledge, this is the first experimental proof-of-concept for real-time transmission of \ac{HD} video signals over optical wireless channels using \ac{SDI} interfaces, supporting \ac{FHD} and 4K \ac{UHD} resolutions at data rates up to $6$~Gb/s, while achieving an ultra-low latency of less than $35$~ns for the end-to-end system.

The rest of the paper is organized as follows. In Section~\ref{Sec:SDI-over-LiFi}, the design and implementation of the \ac{SDI} over \ac{LiFi} system are elaborated. In Section~\ref{Sec:Results}, measurement results are presented and analyzed in terms of the real-time eye diagram and end-to-end latency. In Section~\ref{Sec:Conclusions}, concluding remarks are given.

\begin{table}[!t]
\centering
\caption{SMPTE SDI Standards \cite{SMPTE259,SMPTE424,SMPTE2081,SMPTE2082,SMPTE2083}}
\begin{threeparttable}
\begin{tabular}{l|l|l|l}
\textbf{Variant} & \textbf{Standard}    & \textbf{Data Rate} & \textbf{Video Format}      \\
\hline
SD-SDI               & ST 259         & $270$~Mb/s           & SD (480i @ 30 fps)         \\
SD-SDI               & ST 259         & $270$~Mb/s           & SD (576i @ 25 fps)         \\
HD-SDI               & ST 292         & $1.485$~Gb/s         & HD (720p @ 30 fps)         \\
HD-SDI               & ST 292         & $1.485$~Gb/s         & HD (1080i @ 25 fps)        \\
3G-SDI               & ST 424         & $2.97$~Gb/s          & FHD (1080p @ 50 fps)   \\
3G-SDI               & ST 424         & $2.97$~Gb/s          & FHD (1080p @ 60 fps)   \\
6G-SDI               & ST 2081        & $5.94$~Gb/s          & 4K UHD (2160p @ 30 fps)    \\
12G-SDI              & ST 2082        & $11.88$~Gb/s         & 4K UHD (2160p @ 60 fps)    \\
24G-SDI              & ST 2083        & $23.76$~Gb/s         & 4K UHD (2160p @ 120 fps)   \\
24G-SDI              & ST 2083        & $23.76$~Gb/s         & 8K UHD (4320p @ 30 fps)    \\
\hline
\end{tabular}
    \begin{tablenotes}
        \item[] fps denotes the frame rate unit [frame/s].
    \end{tablenotes}
    \end{threeparttable}
\label{Tab:SDI_Variants}
\vspace{-5pt}
\end{table}

\begin{figure}[!t]
	\centering
	\includegraphics[width=0.8\linewidth]{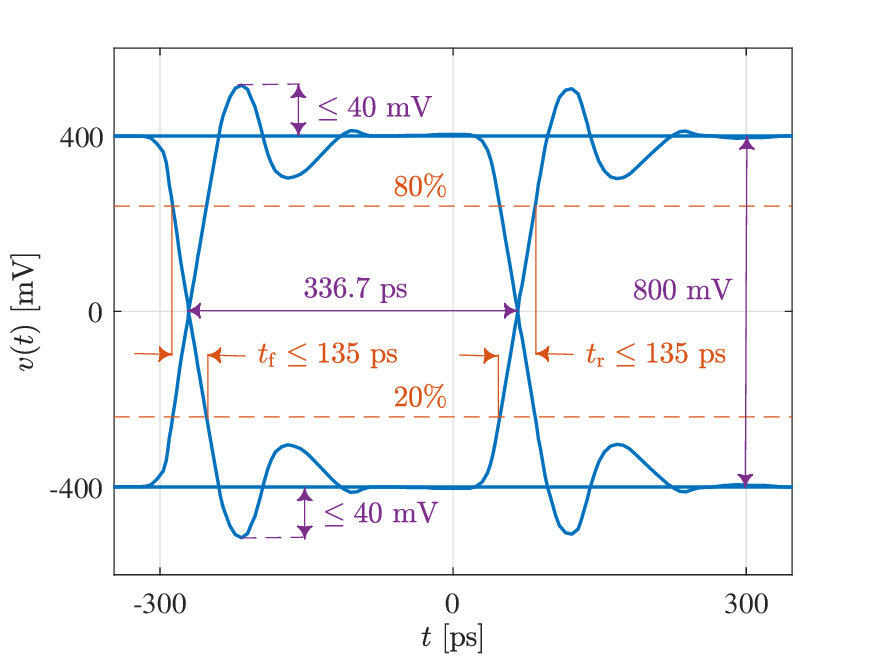}
	\caption{3G-SDI waveform measurement dimensions \cite{SMPTE424}.}
    \label{Fig:3G_SDI_Eye_Parameters}
\end{figure}

\section{SDI over LiFi System Design} \label{Sec:SDI-over-LiFi}
In this section, we present the design of the proposed \ac{SDI}-over-{LiFi} system. First, we provide an overview of the key specifications of the \ac{SDI} signal interface with the objective of determining the dynamic range and bandwidth requirements for the system. Then, we present the experimental setup of the \ac{VCSEL}-based optical wireless video transmission link. Table~\ref{Tab:SDI_Variants} provides the specifications of various \ac{SMPTE} \ac{SDI} standards in terms of their supported data rates and video formats \cite{SMPTE259,SMPTE424,SMPTE2081,SMPTE2082,SMPTE2083}.

\newcounter{mycounter1}
\setcounter{mycounter1}{\value{figure}}
\setcounter{figure}{2}
\begin{figure*}[!t]
    \centering
    \includegraphics[width=\linewidth]{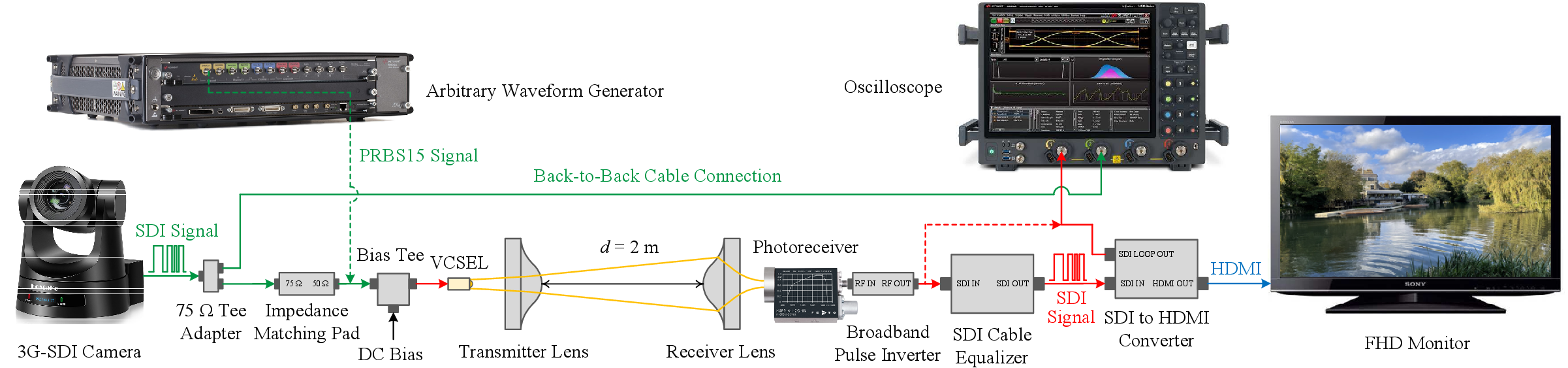}
    \caption{Experimental setup for the proposed VCSEL-based SDI-over-LiFi link. The dashed green connection is used for emulating 3G-SDI signal transmission. The dashed red connection from the receiver output is used for latency measurements relative to the back-to-back signal from the camera.}
    \label{Fig:Experimental_Setup}
\end{figure*}
\setcounter{figure}{\value{mycounter1}}

\subsection{3G-SDI Signal Interface} \label{Subsec:3G-SDI}
We use an \ac{HD} camera for live video streaming. The camera is equipped with a 3G-\ac{SDI} interface that generates a bit stream at a nominal bit rate of $3$~Gb/s. According to the \ac{SMPTE} ST 424 standard for $3$~Gb/s serial interface \cite{SMPTE424}, the 3G-\ac{SDI} signal operates exactly at $2.97$~Gb/s, and it has a rectangular pulse shape with a peak-to-peak amplitude of $800$~mV $\pm~10\%$ and a \ac{DC} offset of $0.0$~V $\pm~0.5$~V. Therefore, it is a bipolar signal that switches between $\pm 400$~mV values. Fig.~\ref{Fig:3G_SDI_Eye_Parameters} shows the eye diagram specifications for the 3G-\ac{SDI} waveform as defined by the \ac{SMPTE} ST 424 standard \cite{SMPTE424}. The duration of one \ac{UI} is $336.7$~ps, corresponding to $2.97$~Gb/s. The rise and fall times of the waveform, denoted by $t_\mathrm{r}$ and $t_\mathrm{f}$, respectively, are determined between the $20\%$ and $80\%$ points of the peak-to-peak amplitude. In this case, $t_\mathrm{r}\leq135$~ps and $t_\mathrm{f}\leq135$~ps such that $\lvert t_\mathrm{r}-t_\mathrm{f}\rvert\leq50$~ps. The waveform exhibits an overshoot and an undershoot of up to $40$~mV, equivalent to $10\%$ of the peak amplitude.

In \ac{SDI} video signals, binary information bits are encoded using a scrambled \ac{NRZI} channel coding scheme \cite[Annex A]{SMPTE424}. Unlike \ac{NRZ} encoding where binary data bits of `1' and `0' are represented by rectangular pulses of polarities $A$ and $-A$, respectively, in \ac{NRZI} encoding, transitions between the two amplitude levels occur only when a `1' is transmitted, otherwise the amplitude level remains unchanged \cite{Proakis2008DigitalCommunications}. As a result, \ac{NRZI} is a differential encoding scheme followed by \ac{NRZ} signaling. A key advantage of differential encoding is the ability to detect errors in the event that some expected changes in the state of the signal are missing. This helps to maintain data synchronization even in scenarios where a long sequence of the same binary data is present, allowing the \ac{SDI} signal to be properly synchronized at the receiver end. Moreover, according to the Nyquist-Shannon theorem, for baseband binary signal transmission at a bit rate of $R_\mathrm{b}$, a system bandwidth of at least $B=0.5R_\mathrm{b}$ is required \cite{Haykin2009CommunicationSystems}. Thus, the minimum bandwidth requirement for 3G-\ac{SDI} signal is $1.5$~GHz or $1.485$~GHz to be precise. For safety margin, we may consider adding headroom for filtering and equalization tolerance, and design the front-end components of the system for a higher bandwidth of $B=0.7R_\mathrm{b}$, if needed. In this case, the minimum bandwidth requirement increases to $2$~GHz.

\subsection{Experimental Setup}
\begin{figure}[!t]
	\centering
	\includegraphics[width=\linewidth]{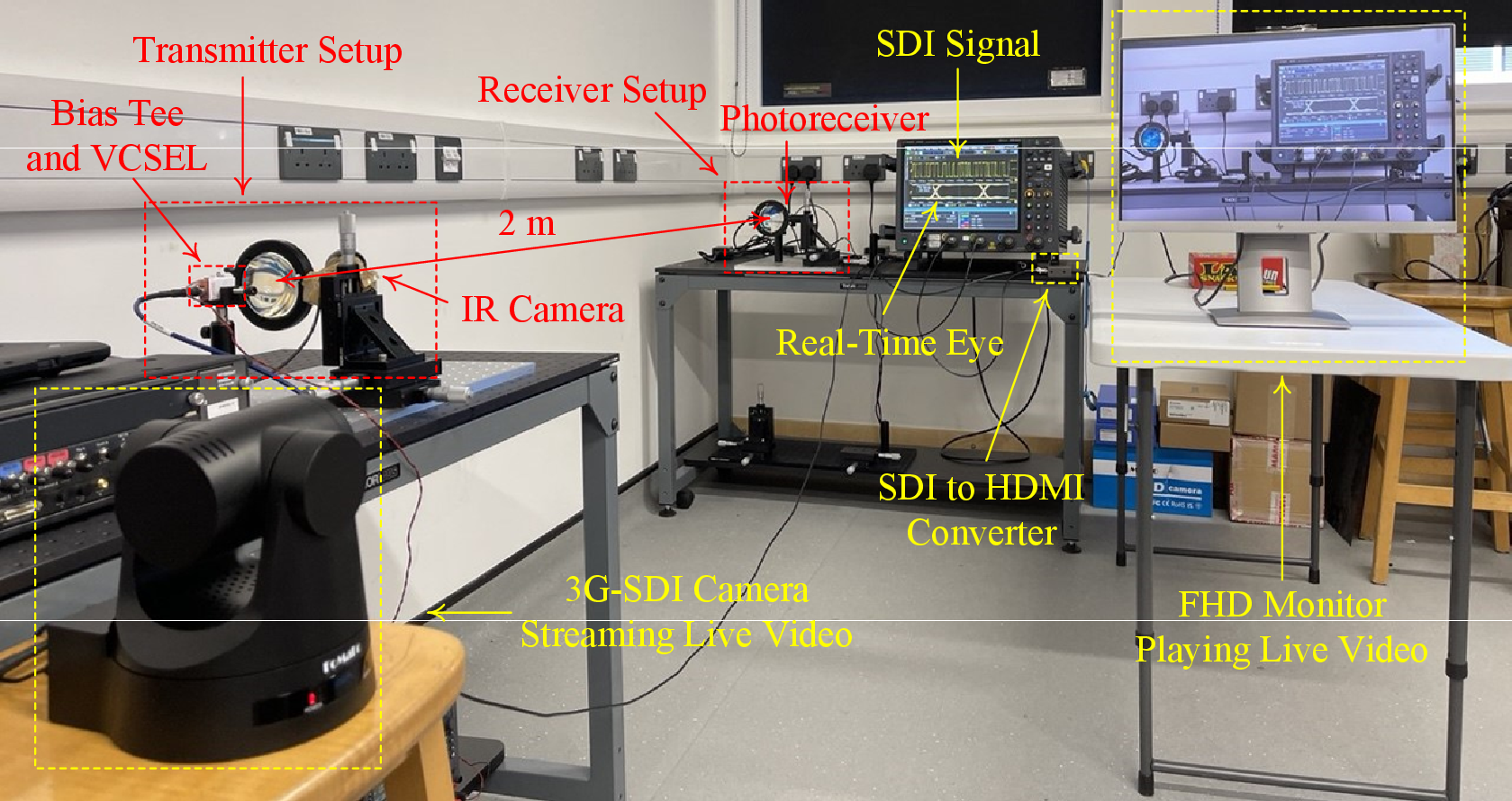}
	\caption{Live video streaming from a 3G-SDI camera to a FHD monitor via the VCSEL-based SDI-over-LiFi link over a $2$~m distance.}
    \label{Fig:Live_Demo}
\end{figure}

Fig.~\ref{Fig:Live_Demo} shows the experimental setup of the proposed real-time \ac{SDI} over \ac{LiFi} system, streaming live video from an \ac{HD} camera to a \ac{FHD} monitor over an optical wireless link of length $2$~m. Fig.~\ref{Fig:Experimental_Setup} shows the complete experimental setup for the end-to-end system. The components used in the system setup are listed in Table~\ref{Tab:System_Components}. The 3G-\ac{SDI} signal is available at the \ac{SDI} output port of the camera via a \ac{BNC} type coaxial connector. This wideband analog signal is directly modulated on the intensity of a $940$~nm \ac{SM} \ac{VCSEL} at the transmitter by using a \ac{BT}. The \ac{VCSEL} was developed, fabricated and packaged by \ac{VIS} GmbH, and underwent wafer-level testing prior to full characterization for high-speed optical communications \cite{NLedentsov2022Advances}. It has a \ac{MA} structure consisting of four \acp{VCSEL} that are arranged in a compact $2\times2$ mini-array configuration with a pitch of $<15$~nm. These four \acp{VCSEL} are driven by a common pair of contacts and their outputs are strongly coupled such that they effectively operate as a single \ac{VCSEL} \cite{NLedentsov2022Advances}. In this way, the output power increases while preserving the \ac{SM} emission spectrum. The \ac{MA} \ac{VCSEL} exhibits a near-Gaussian beam profile with a \ac{FF} \ac{FWHM} divergence angle of $18^\circ$, a peak optical power of $14$~mW, and a $3$-dB modulation bandwidth of $18$~GHz. It also offers a linear input dynamic range between a forward threshold current of $2$~mA and a roll-over current of $30$~mA.

\begin{table}[t!]
	\centering
	\caption{System Components}
    \begin{threeparttable}
	\begin{tabular}{l|l|l|l}
		\textbf{Component}    & \textbf{Manufacturer}   & \textbf{Model}   & \textbf{Bandwidth}\hspace{1pt}\textbf{/}\hspace{1pt}\textbf{Rate}            \\ \hline
		Camera       & FoMaKo         & PTZ 3G-SDI                    & $3$~Gb/s             \\
        RFS          & Telegartner    & $75$~{\textOmega} Tee Adapter & $4$~GHz              \\
        BT           & Mini-Circuits  & ZX85-12G-S+                   & $12$~GHz             \\
        ACL          & Thorlabs       & ACL7560U-B                    & --                   \\
        PR           & FEMTO          & HSPR-X-I-2G-IN                & $2$~GHz              \\
        CEQ          & Extron         & 12G HD-SDI 101                & $12$~Gb/s            \\
        SHC          & Blackmagic     & 12G                           & $12$~Gb/s            \\
        AWG          & Keysight       & M8195A                        & $6$~GHz, $16$~GSa/s  \\
        OSC          & Keysight       & UXR0104B                      & $10$~GHz, $128$~Ga/s \\ \hline
	\end{tabular}
    \begin{tablenotes}
        \item[] RFS: RF Splitter; PR: Photoreceiver; SHC: SDI to HDMI Converter
    \end{tablenotes}
    \end{threeparttable}
	\label{Tab:System_Components}
\end{table}

In optical \ac{IM{/}DD} systems, the modulating signal is non-negative and real-valued. In particular, in laser-based optical \ac{IM{/}DD} systems using \ac{NRZ{-}OOK} modulation scheme, a \ac{DC} offset is added to the bipolar \ac{NRZ} signal. To this end, the laser is typically biased slightly below its turn-on threshold, allowing it to produce a continuous optical signal during the transmission period for a data bit of `1' and to be effectively turned off for a data bit of `0' \cite{Agrawal2005LightwaveTechnology}. This is a viable approach as long as the laser source can be switched on and off at a speed matching the bit rate of the \ac{NRZ} pulse. Although the modulation bandwidth of the \ac{VCSEL} is larger than the bit rate of the 3G-\ac{SDI} signal, we opt for a sufficiently high \ac{DC} bias level to ensure operation within the linear dynamic range of the \ac{VCSEL} to further improve the laser response time and avoid nonlinear distortion effects. Based on the \ac{LIV} characteristics of the \ac{VCSEL}, we choose a \ac{DC} operating point in the middle of the linear dynamic range by setting $V_\mathrm{DC}=2.4$~V and $I_\mathrm{DC}=8.42$~mA. Consequently, the \ac{VCSEL} is never turned off and a maximally linear dynamic range is provided for the \ac{SDI} signal swing. The \ac{BT} has a wideband frequency range of $200$~kHz to $12$~GHz with a low \ac{IL} of $\mathrm{IL}<1$~dB over the signal bandwidth. The \ac{SDI} output of the camera is connected to the \ac{BT} using a 12G-\ac{SDI} cable, which supports bit rates up to $12$~Gb/s. \ac{SDI} cables are $75$~{\textOmega} wideband coaxial cables with \ac{BNC} connectors\footnote{Compared to the characteristic impedance of $50$~{\textOmega}, commonly used in \acs{RF} test and measurement equipment, $75$~{\textOmega} coaxial cables with \ac{BNC} connectors used for analog video transmission exhibit lower attenuation \cite{Fischer2020DigitalVideo}.}. Since the \ac{BT} has an input impedance of $50$~{\textOmega}, we use an impedance matching pad to convert $75$~{\textOmega} to $50$~{\textOmega} before feeding the \ac{SDI} signal into the \ac{BT}, as shown in Fig.~\ref{Fig:Experimental_Setup}. The biased \ac{SDI} signal is then fed into the \ac{VCSEL} using a $26$~GHz \ac{RF} coaxial cable with $\mathrm{IL}<1$~dB over the desired bandwidth. To achieve beam collimation at the transmitter, an \ac{ACL} is positioned in front of the \ac{VCSEL}. This \ac{ACL} has a diameter of $50$~mm, a focal length of $40$~mm, a \ac{NA} of $0.6$, and is equipped with an \ac{ARC} for $650$--$1050$~nm wavelength range.

At the receiver side, located $2$~m away from the transmitter, the incident light beam is collected by an \ac{ACL}, identical to the transmitter lens, and is focused on a high-speed photodetector. Specifically, we employ a wideband \ac{PIN} photoreceiver with lower and upper $-3$~dB cut-off frequencies of $10$~kHz and $2$~GHz, respectively. It consists of an InGaAs \ac{PD} with an active area of $100$~{\textmu}m, followed by an integrated low-noise \ac{TIA}. The \ac{PD} operates over a spectral range of $900$--$1700$~nm with a maximum responsivity of $0.95$~A/W at $1550$~nm, and it has a responsivity of about $0.5$~A/W at the \ac{VCSEL} wavelength (i.e., $940$~nm). The \ac{TIA} is an inverting amplifier that provides a transimpedance gain of $5.0\times10^{3}$~V/A, resulting in a conversion gain of $2.5\times10^{3}$~V/W at $940$~nm. The photoreceiver generates an \ac{AC}-coupled output voltage with a maximum peak-to-peak amplitude of $2.0$~V. The output signal has an inverted polarity with respect to the input \ac{SDI} signal. To revert back to the original polarity, we employ a \ac{BPI} at the receiver output, as shown in Fig.~\ref{Fig:Experimental_Setup}. The \ac{BPI} is a two-port device that introduces a phase shift of $180^\circ$ relative to its input signal while maintaining a flat group delay over a broadband frequency range of $1$~MHz to $40$~GHz. It has an \ac{IL} of about $2$~dB over the \ac{SDI} signal bandwidth. The detected signal is subsequently fed into an \ac{SDI} \ac{CEQ} using a 12G-\ac{SDI} cable. The \ac{CEQ} automatically adapts to \ac{SMPTE} serial digital video standards for \ac{SDI} signals up to a 12G-\ac{SDI} data rate. It compensates for cable attenuation of up to $-30$~dB over distances of up to $240$~m, specifically for \ac{HD}-\ac{SDI} signals. It also reshapes and restores signal timing using an embedded reclocking repeater, eliminating high-frequency jitter\footnote{The reclocking repeater does not decode, descramble, or deserialize the data stream. Rather, it equalizes, slices, and reclocks the input signal using a \ac{PLL} to reject jitter and noise, thereby recovering a clean binary signal, which can then be forwarded to an \ac{SDI} line driver \cite{Watkinson2004DigitalInterface}.}. We employ an \ac{SDI} to \ac{HDMI} converter to convert the received \ac{SDI} video stream to an \ac{HDMI} signal and cast it on the \ac{FHD} monitor. This converter also has an \ac{SDI} loop out, providing a replica of its input \ac{SDI} signal, which is connected to the \ac{OSC} for real-time eye diagram measurements, as shown in Fig.~\ref{Fig:Experimental_Setup}. The \ac{OSC} performs \ac{ADC} at a sampling rate of $32$~GSa/s to improve measurement accuracy and capture fine details of the received 3G-\ac{SDI} signal.

Before connecting the camera and monitor, proper operation of the \ac{SDI} over \ac{LiFi} system should be verified. To this end, we generate a \ac{PRBS} of length $2^{15}-1$ bits in MATLAB. Then, by using an \ac{AWG}, we turn the resulting \ac{PRBS}15 bit stream into an \ac{NRZ} encoded waveform with the same bit rate as the 3G-\ac{SDI} signal based on \ac{BPSK} modulation with rectangular pulse shaping. After \ac{DC} biasing, the \ac{PRBS}15 \ac{NRZ} signal is sent over the \ac{VCSEL}-based optical wireless link in order to emulate 3G-\ac{SDI} signal transmission. This is indicated by a dashed green connection in Fig.~\ref{Fig:Experimental_Setup}. The real-time eye diagram of the equalized signal at the receiver yields an accurate estimate of the system performance during 3G-\ac{SDI} signal transmission. Eye diagram measurements are discussed in the following section.

\section{Measurement Results} \label{Sec:Results}
We present experimental results for the real-time eye pattern and end-to-end latency measurements. First, let us introduce an important quality metric used for performance analysis.

\subsection{Link Quality Metric} \label{Quality_Metric}
A widely used performance metric for optical signals is the $Q$-factor \cite{Freude2012QualityMetrics}. For an \ac{IM{/}DD} optical communication system based on \ac{NRZ} \ac{OOK} modulation, the $Q$-factor is defined as \cite{Freude2012QualityMetrics}:
\begin{equation}
    Q = \frac{V_\mathrm{s}}{\sigma_\mathrm{n}},
    \label{Eq:Q-factor}
\end{equation}
where $V_\mathrm{s}$ is the amplitude of the received signal and $\sigma_\mathrm{n}$ is the standard deviation of noise at the receiver. The corresponding \ac{SNR} and \ac{BER} are given by \cite{Freude2012QualityMetrics}:
\begin{equation}
    \mathrm{SNR} = Q^2,
    \label{Eq:SNR}
\end{equation}
\begin{equation}
    \mathrm{BER} = \frac{1}{2}\mathrm{erfc}\left(\frac{Q}{\sqrt{2}}\right),
    \label{Eq:BER}
\end{equation}
where $\mathrm{erfc}$ denotes the complementary error function given by $\mathrm{erfc}(z)=\frac{2}{\sqrt{\pi}}\int_{z}^{\infty}e^{-u^2}du$.

Since \ac{SDI} is a point-to-point interface, the feasibility of an \ac{SDI} link to operate reliably at a given data rate is determined primarily by the link distance (i.e., cable length). In practice, the \ac{BER} performance rapidly deteriorates beyond a critical cable length known as the crash knee \cite{Watkinson2004DigitalInterface}. This critical distance is identified at the point where at most $1$ bit error occurs in a given period of time. In particular, \ac{HD}-\ac{SDI} links with a target performance of $1$ bit error per second and $1$ bit error per hour, corresponding to \ac{BER} thresholds of $4.7\times10^{-9}$ and $1.3\times10^{-12}$, can reach distances of up to $188$~m and $182$~m, respectively, based on the \ac{PAL} \ac{TV} broadcasting system \cite{Watkinson2004DigitalInterface}. From \eqref{Eq:SNR} and \eqref{Eq:BER}, these \ac{BER} values translate into a minimum $Q$-factor of $5.7$ and $7$, equivalent to a minimum required \ac{SNR} of $15.2$~dB and $16.9$~dB, respectively.

\subsection{Real-Time Eye Pattern} \label{Eye_Pattern}
The eye pattern is an effective tool for evaluating the quality of signals in digital communications. An eye pattern is formed by the synchronized superposition of all possible realizations of the signal of interest when observed within a given signaling interval \cite{Haykin2009CommunicationSystems}. The vertical and horizontal eye openings are key characteristics of the eye diagram. The vertical eye opening reflects the noise margin of the signal and is indicative of the \ac{SNR}, while the horizontal eye opening corresponds to the timing margin and provides insight into the extent of timing jitter and \ac{ISI} \cite{Haykin2009CommunicationSystems}.

\addtocounter{figure}{1}
\begin{figure}[!t]
    \centering
    \subfloat[At $2.97$~Gb/s for 3G-SDI \label{Fig:EyeDiagram_PRBS14_a}]{\includegraphics[width=0.7\linewidth, keepaspectratio=true]{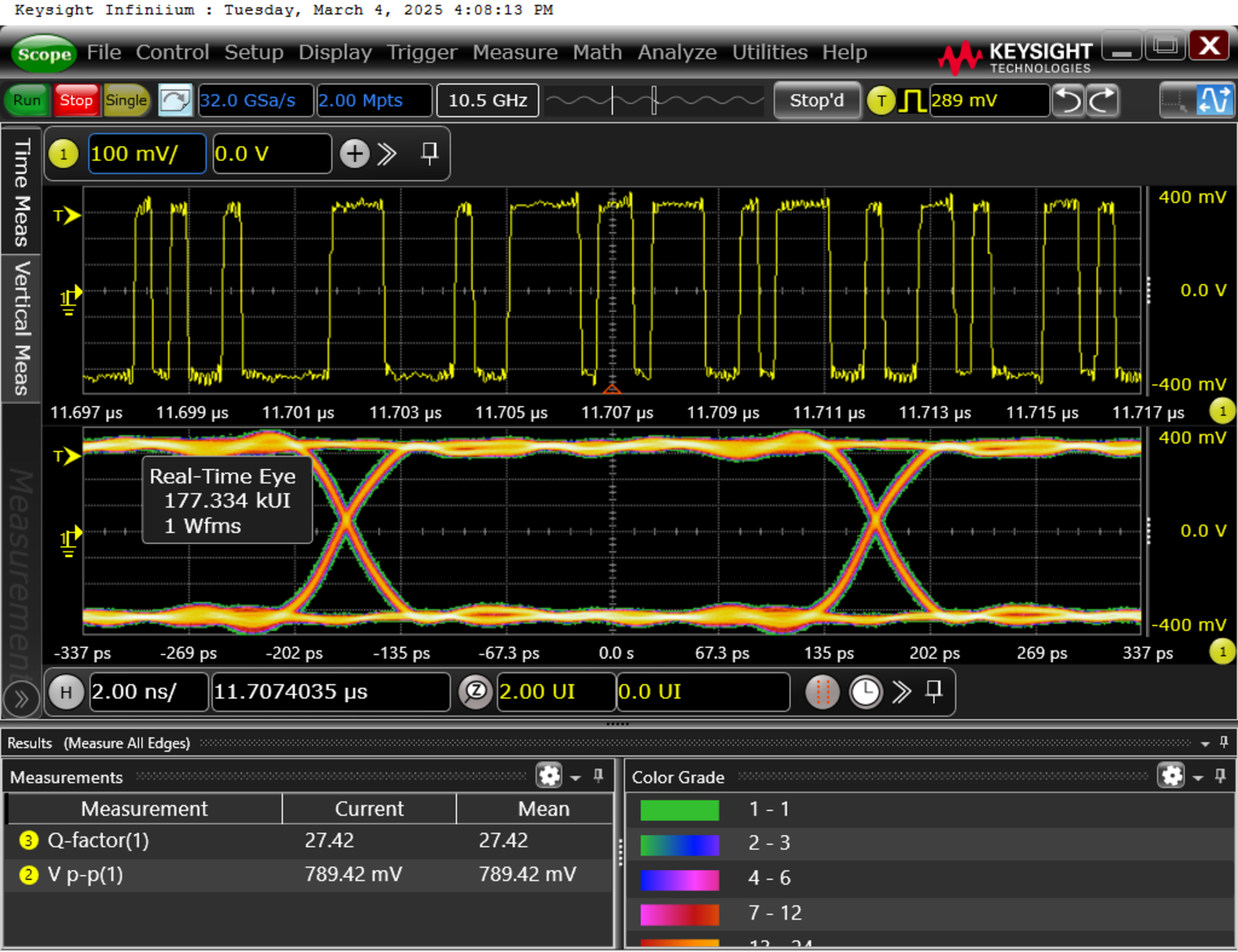}} \\
    \subfloat[At $5.94$~Gb/s for 6G-SDI \label{Fig:EyeDiagram_PRBS14_b}]{\includegraphics[width=0.7\linewidth, keepaspectratio=true]{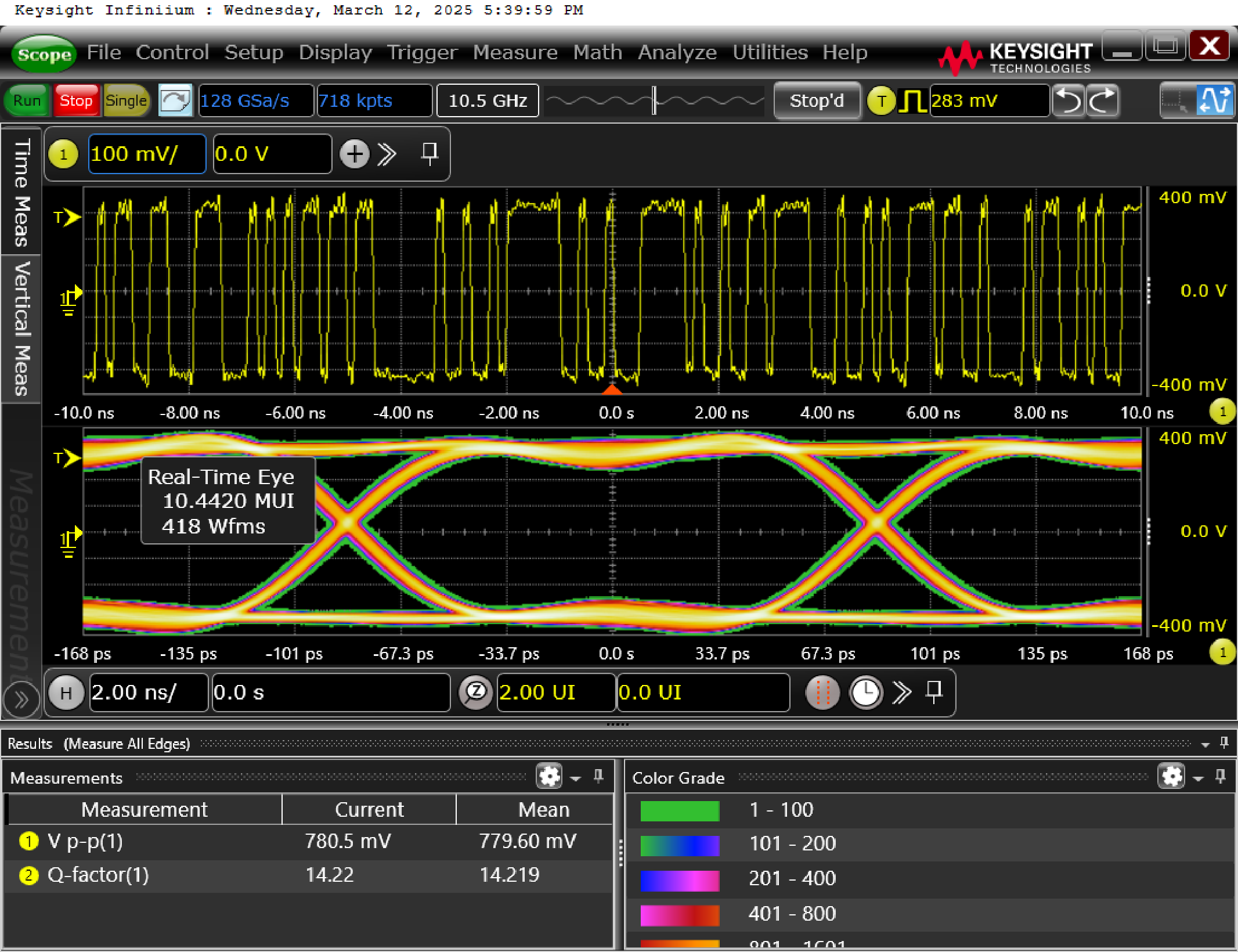}}
    \caption{Real-time eye patterns for received PRBS15 NRZ test signals.}
    \label{Fig:EyeDiagram_PRBS14}
\end{figure}

Fig.~\ref{Fig:EyeDiagram_PRBS14} shows the real-time eye patterns on the \ac{OSC} for received \ac{PRBS}15 \ac{NRZ} signals at data rates of $2.97$~Gb/s and $5.94$~Gb/s to emulate 3G-\ac{SDI} and 6G-\ac{SDI} video transmissions, respectively, as indicated in Table~\ref{Tab:SDI_Variants}. It can be seen that the eye diagrams in both cases are wide-open both horizontally and vertically, demonstrating the excellent performance of the \ac{SDI} over \ac{LiFi} system. For the \ac{PRBS}15 \ac{NRZ} test link at $2.97$~Gb/s, as shown in Fig.~\subref*{Fig:EyeDiagram_PRBS14_a}, a $Q$-factor of $27.42$ is achieved, which translate into an \ac{SNR} value of $28.76$~dB. When the data rate is doubled, as shown in Fig.~\subref*{Fig:EyeDiagram_PRBS14_b}, the $Q$-factor reduces to $14.22$, equivalent to an \ac{SNR} value of $23.06$~dB. However, this $Q$-factor is more than twice greater than the $Q$-factor threshold of $7$ required to keep the \ac{BER} level below $1.3\times10^{-12}$. This evinces that the system already supports the 6G-\ac{SDI} signal transmission with almost error-free performance.

\begin{figure}[!t]
    \centering
    \includegraphics[width=0.7\linewidth, keepaspectratio=true]{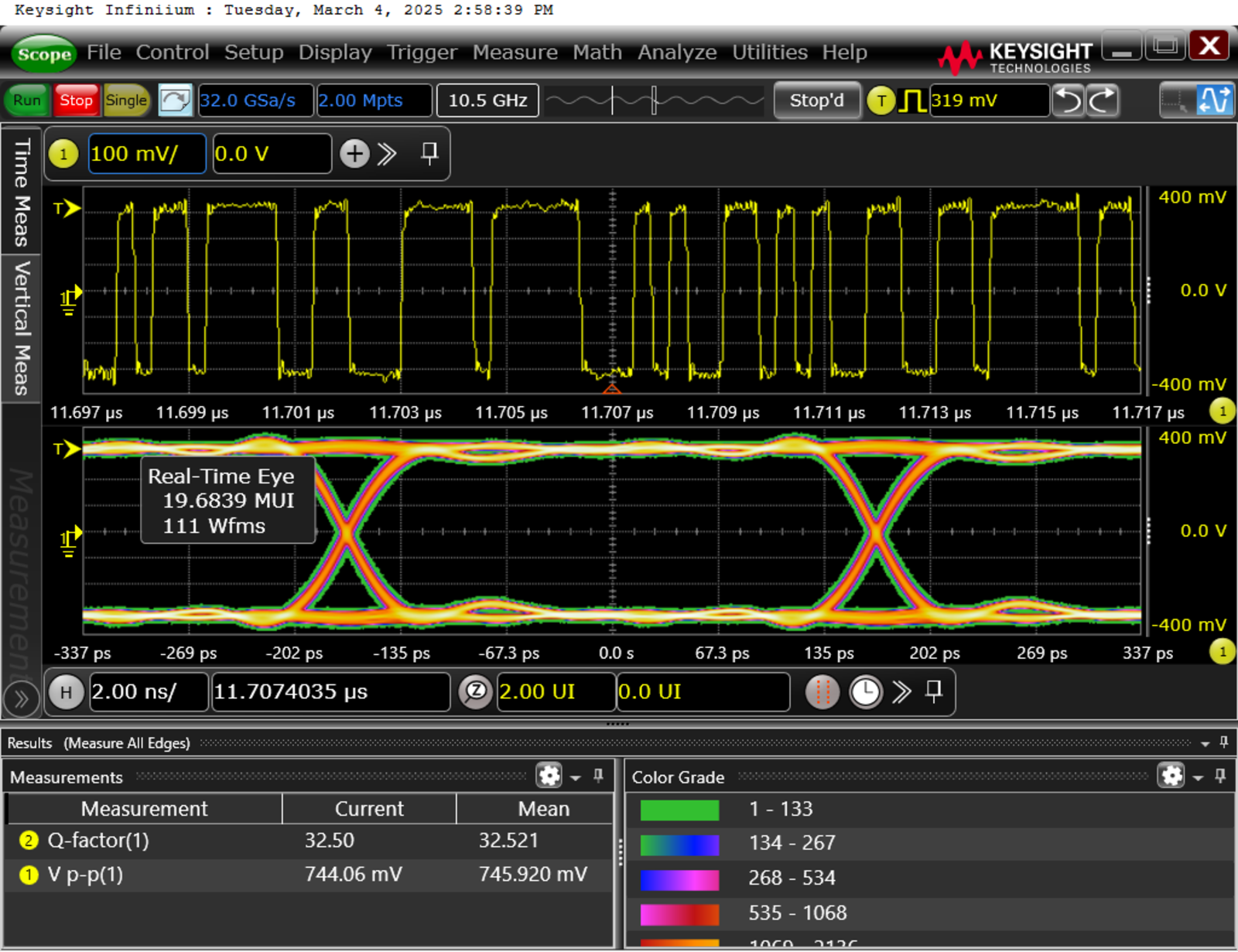}
    \caption{Real-time eye pattern for the received 3G-SDI signal from the camera.}
    \label{Fig:EyeDiagram_RealTime}
\end{figure}

Fig.~\ref{Fig:EyeDiagram_RealTime} shows the real-time eye diagram on the \ac{OSC} for the received 3G-\ac{SDI} signal from the camera. It can be observed that the $Q$-factor is improved to $32.50$ compared to the corresponding \ac{PRBS}15 \ac{NRZ} test link, as shown in Fig.~\subref*{Fig:EyeDiagram_PRBS14_a}, resulting in an \ac{SNR} value of $30.24$~dB. This is due to the fact the \ac{CEQ} has optimum performance for standard \ac{SDI} video signals. With this \ac{SNR} level, the \ac{SDI} over \ac{LiFi} link operates at an extremely low \ac{BER} based on \eqref{Eq:BER}.

\subsection{End-to-End Latency}
The system setup used for end-to-end latency measurements is shown in Fig.~\ref{Fig:Experimental_Setup}. A replica of the input \ac{SDI} signal is used as the reference signal. To this end, a back-to-back connection is extended from the camera to the \ac{OSC} using a $75$~{\textOmega} \ac{RF} tee adapter (i.e., splitter) at the 3G-\ac{SDI} output of the camera. The tee adapter equally divides the \ac{RF} power between its two output ports\footnote{Although this leads to a $3$~dB loss for each signal path and the \ac{SDI} signal that reaches the \ac{VCSEL} is slightly attenuated, the peak-to-peak amplitude of the signal is still sufficient to ensure a high \ac{SNR}, as reported in Section~\ref{Eye_Pattern}.} over a frequency range of \ac{DC} to $4$~GHz. For time delay estimation, we apply the cross-correlation technique using a large number of signal samples \cite{Jacovitti1993DelayEstimation}. Let $x(t)$ and $y(t)$, respectively, denote the back-to-back signal and the received signal before the \ac{CEQ}, as shown in Fig.~\ref{Fig:Experimental_Setup}. For these two real signals, the sample cross-correlation function is computed by \cite{Jacovitti1993DelayEstimation}:
\begin{equation}
\begin{aligned}
    \hat{R}_{xy}(\tau) &= \frac{1}{N}\sum_{k=1}^{N}x(t)y(t+\tau) \big{|}_{t=kT}, \\
                       &= \frac{1}{N}\sum_{k=1}^{N}x(kT)y(kT+\tau), 
\end{aligned}
\label{Eq:DirectCorrelator}
\end{equation}
where $N$ is the total number of samples, and $T$ is the sampling period, and $W=(N-1)T$ is the duration of the estimation window. The time delay between the two signals, denoted by $\tau_\mathrm{d}$, is determined by the point in time where the direct correlator in \eqref{Eq:DirectCorrelator} is maximized, indicating the relative time shift at which they align best with each other in terms of similarity. This can be expressed as \cite{Jacovitti1993DelayEstimation}:
\begin{equation}
    \tau_\mathrm{d} = \operatorname*{arg\;max}_{\tau} \ \hat{R}_{xy}(\tau).
    \label{Eq:TimeDelay}
\end{equation}
Note that this is the relative delay between the output signal and the reference signal through the back-to-back connection. Let $\tau_\mathrm{ow}$ and $\tau_\mathrm{bb}$ denote the end-to-end latency of the \ac{SDI} over optical wireless link and the delay caused by the back-to-back cable, respectively. We have $\tau_\mathrm{d}=\tau_\mathrm{ow}-\tau_\mathrm{bb}$, which leads to:
\begin{equation}
    \tau_\mathrm{ow} = \tau_\mathrm{d}+\tau_\mathrm{bb}.
    \label{Eq:Latency}
\end{equation}

\begin{figure}[!t]
    \centering
    \subfloat[Normalized cross-correlation function \label{Fig:CrossCorrelation_a}]{\includegraphics[width=0.8\linewidth, keepaspectratio=true]{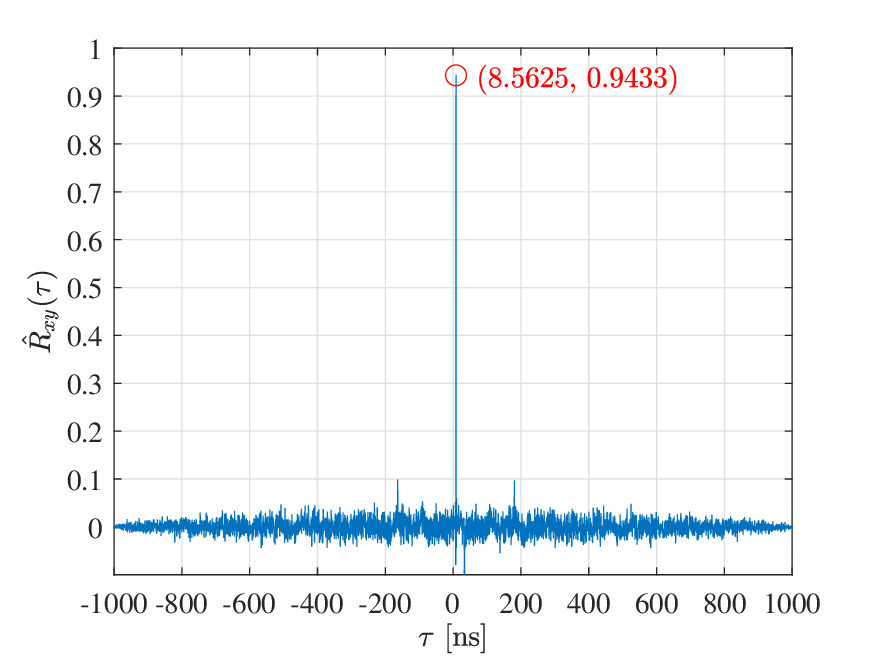}} \\
    \vspace{-10pt}
    \subfloat[$x(t)$ vs. $y(t+\tau_\mathrm{d})$ for $\tau_\mathrm{d}=8.56$~ns \label{Fig:CrossCorrelation_b}]{\includegraphics[width=0.8\linewidth, keepaspectratio=true]{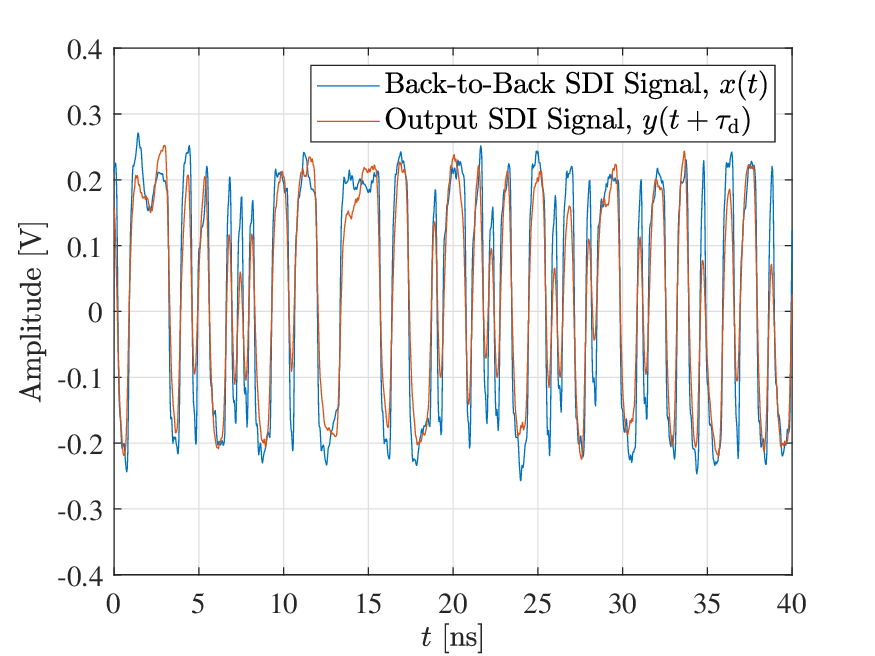}}
    \caption{Estimating the time delay between the input SDI signal, $x(t)$, and the output SDI signal, $y(t)$, based on their cross-correlation, $\hat{R}_{xy}(\tau)$.}
    \label{Fig:CrossCorrelation}
\end{figure}

Fig.~\ref{Fig:CrossCorrelation} shows the time delay estimation based on maximizing the cross-correlation between the input and output \ac{SDI} signals. With the \ac{OSC} operating at $32$~GSa/s, for each signal, a total of $N=32001$ samples are obtained over an interval of length $W=1000$~ns to compute the cross-correlation function, as plotted in Fig.~\subref*{Fig:CrossCorrelation_a}, resulting in $\tau_\mathrm{d}=8.56$~ns. To verify this value, a time shift of $t+\tau_\mathrm{d}$ is introduced to the output signal to be compared with the input signal. For clarity, the results are plotted in a $40$~ns interval, as shown Fig.~\subref*{Fig:CrossCorrelation_b}, confirming that the two signals are in perfect agreement when the computed time delay is applied. The delay due to the back-to-back cable is directly measured using a broadband \ac{PNA} based on the the forward voltage gain (i.e., $S_{21}$) of the cable. It is found to be $\tau_\mathrm{bb}=12.18$~ns across the operating bandwidth of the cable (i.e., $>6$~GHz). Based on \eqref{Eq:Latency}, we obtain:
\begin{equation}
    \tau_\mathrm{ow} = 8.56 + 12.18 = 20.74~\text{ns}.
    \label{Eq:End_to_End_Latency}
\end{equation}
The \ac{CEQ} has a flat latency of $14$~ns for \ac{SDI} video equalization and reclocking. Taking into account this delay component, the end-to-end latency of the \ac{SDI} over \ac{LiFi} system is obtained as:
\begin{equation}
    \tau_\mathrm{SDI} = 20.74 + 14 = 34.74~\text{ns}.
    \label{Eq:End_to_End_Latency}
\end{equation}
We note that the \ac{SDI} to \ac{HDMI} converter introduces a delay on the order of a few scan lines of video during signal conversion, which is typically less than one video frame. According to the \ac{SMPTE} ST 424 standard \cite{SMPTE424}, the direct mapping of an \ac{FHD} video stream with a $1080$p resolution (i.e., $1920\times1080$) into an \ac{SDI} data stream using 3G-SDI Level~A format requires a total line count of $1125$ per frame. For a frame rate of $60$~frame/s, the frame duration is $1/60 \approx 16.67$~ms, and thus the line duration is $16.67/1125 \approx 14.81$~{\textmu}s. Consequently, the delay associated with the \ac{SDI} to \ac{HDMI} conversion for $5$ to $10$ lines is estimated as $74$ to $148$~{\textmu}s.

\section{Conclusions} \label{Sec:Conclusions}
In this paper, we have demonstrated real-time and error-free transmission of \ac{HD} video signals over a \ac{VCSEL}-based optical wireless link using a high-performance \ac{SDI} over \ac{LiFi} system. Based on a 3G-\ac{SDI} camera, we have experimentally validated the real-time transmission of a live video stream with a $1080$p \ac{FHD} resolution at $60$~frame/s and a data rate of $2.97$~Gb/s over a link distance of $2$~m with a measured end-to-end latency of $<35$~ns, while achieving an eye diagram $Q$-factor of $>32$, which corresponds to an \ac{SNR} of $>30$~dB. Based on a \ac{PRBS}15 bit stream to emulate 6G-\ac{SDI} signal transmission, the \ac{SDI} over \ac{LiFi} system proves effective in real-time streaming of 4K \ac{UHD} video signals with a resolution of $2160$p at $30$~frame/s and a data rate of $5.94$~Gb/s, with a $Q$-factor of $>14$, which is more than twice the $Q$-factor threshold of $7$ for a \ac{BER} performance of $1.3\times10^{-12}$. Considering the $18$~GHz modulation bandwidth of the \ac{VCSEL}, the $2$~GHz bandwidth of the receiver currently limits the performance of the end-to-end system. This system can potentially support the real-time transmission of 12G-\ac{SDI} or even 24G-\ac{SDI} signals based on a single \ac{VCSEL}, for higher video resolutions up to 8K \ac{UHD}, provided that a receiver with sufficiently high bandwidth is used. Scaling the system design to support higher video resolutions and/or frame rates will be addressed in our future works.

\section*{Acknowledgement}
This work was financially supported by the Engineering and Physical Sciences Research Council (EPSRC) under grant EP/Y037243/1 `Platform Driving The Ultimate Connectivity (TITAN)'. The authors thank Srinjoy Dey for his valuable support in ensuring the successful demonstration of the proposed system during the Connected Futures Festival and the Cambridge 6G Symposium.

\newpage

\bibliographystyle{IEEEtran}
\bibliography{IEEEabrv,references_SDI_over_LiFi}

\begin{IEEEbiography}[{\includegraphics[width=1in,height=1.25in,clip,keepaspectratio]{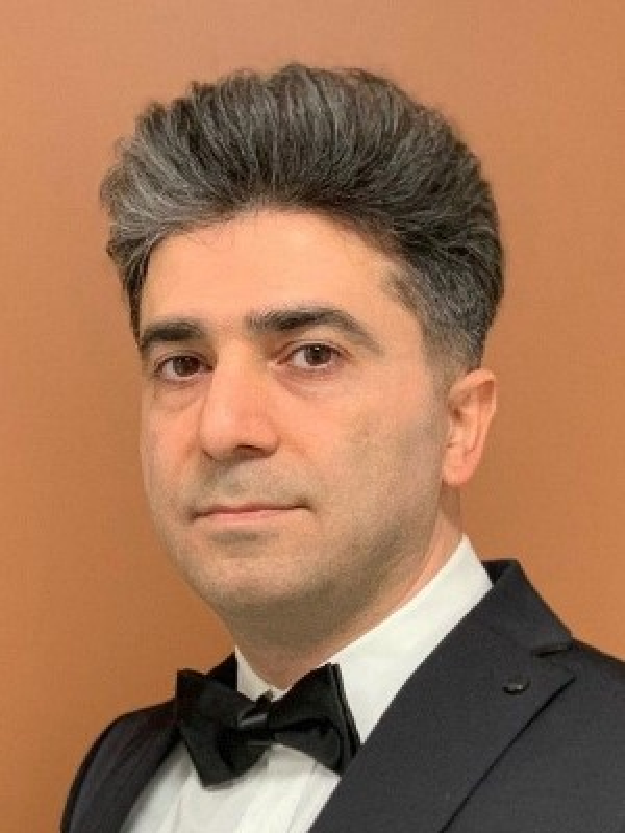}}]{Hossein Kazemi}
(Member, IEEE) received the Ph.D. degree in Electrical Engineering from The University of Edinburgh, U.K., in 2019. He also received the M.Sc. degree in Electrical Engineering (Microelectronic Circuits) from Sharif University of Technology, Tehran, Iran, in 2011, and the M.Sc. degree (Hons.) in Electrical Engineering (Wireless Communications) from Ozyegin University, Istanbul, Turkey, in 2014. He is a Postdoctoral Research Associate at the LiFi Research and Development Center, University of Cambridge, U.K. Dr Kazemi was the recipient of the Best Paper Award for the 2022 IEEE Global Communications Conference (GLOBECOM). His current research interests include the design, analysis and optimization of ultra-high-speed optical wireless communication systems for 6G and beyond networks.
\end{IEEEbiography}


\begin{IEEEbiography}[{\includegraphics[width=1in,height=1.25in,clip,keepaspectratio]{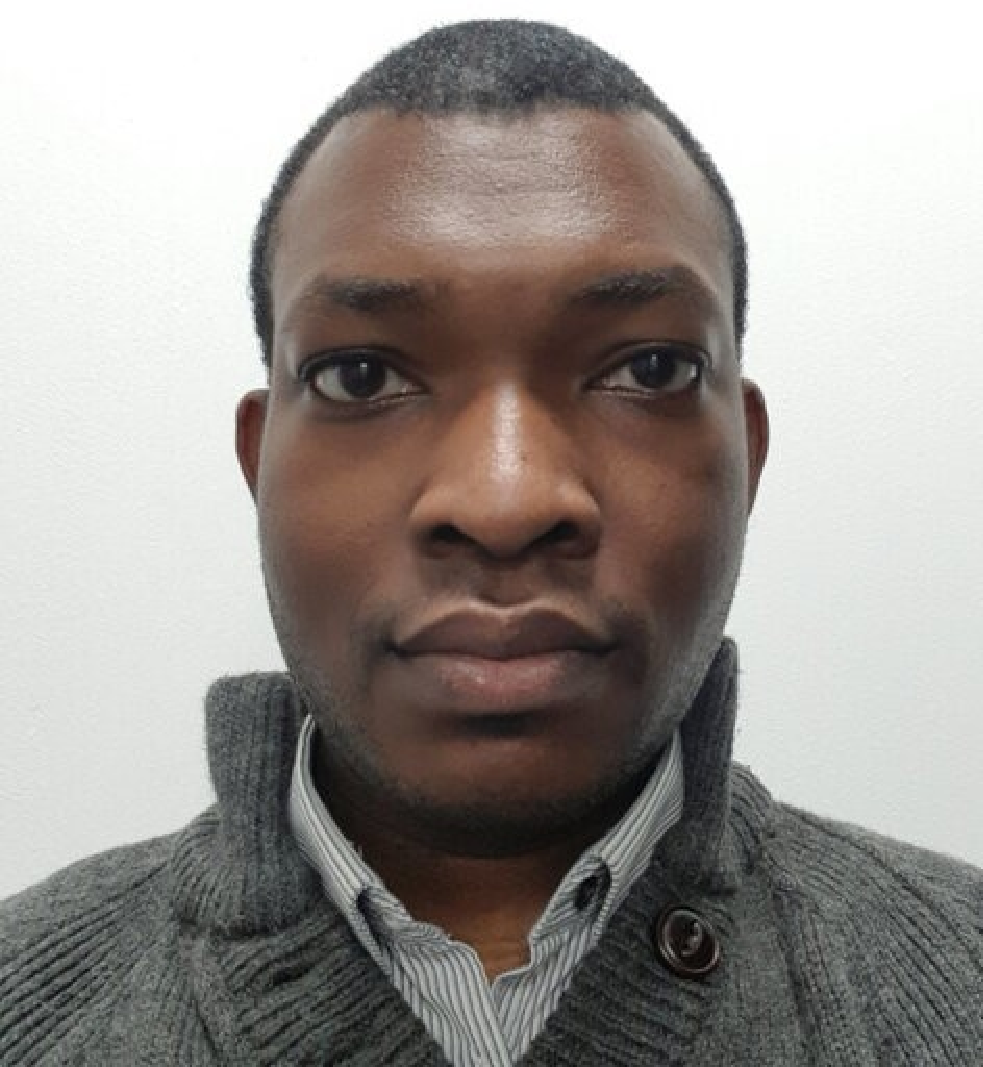}}]{Isaac N. O. Osahon}
(Member, IEEE) received the B.Eng. degree in Electrical and Electronic Engineering (first class hons.) from Covenant University, Ota, Nigeria, in 2012, the M.Sc. degree in Internet Engineering from the University College London, London, in 2015, and the Ph.D. degree at the Institute of Digital Communications, The University of Edinburgh, in 2020. He has worked as a researcher on digital signal processing for optical fiber and wireless communication systems in notable U.K. universities. He is a Postdoctoral Research Associate at the LiFi Research and Development Center, University of Cambridge, U.K. His current research interests include optical communications, advanced modulation schemes, artificial neural networks, digital equalization techniques, visible light positioning and physical layer security.
\end{IEEEbiography}


\begin{IEEEbiography}[{\includegraphics[width=1in,height=1.25in,clip,keepaspectratio]{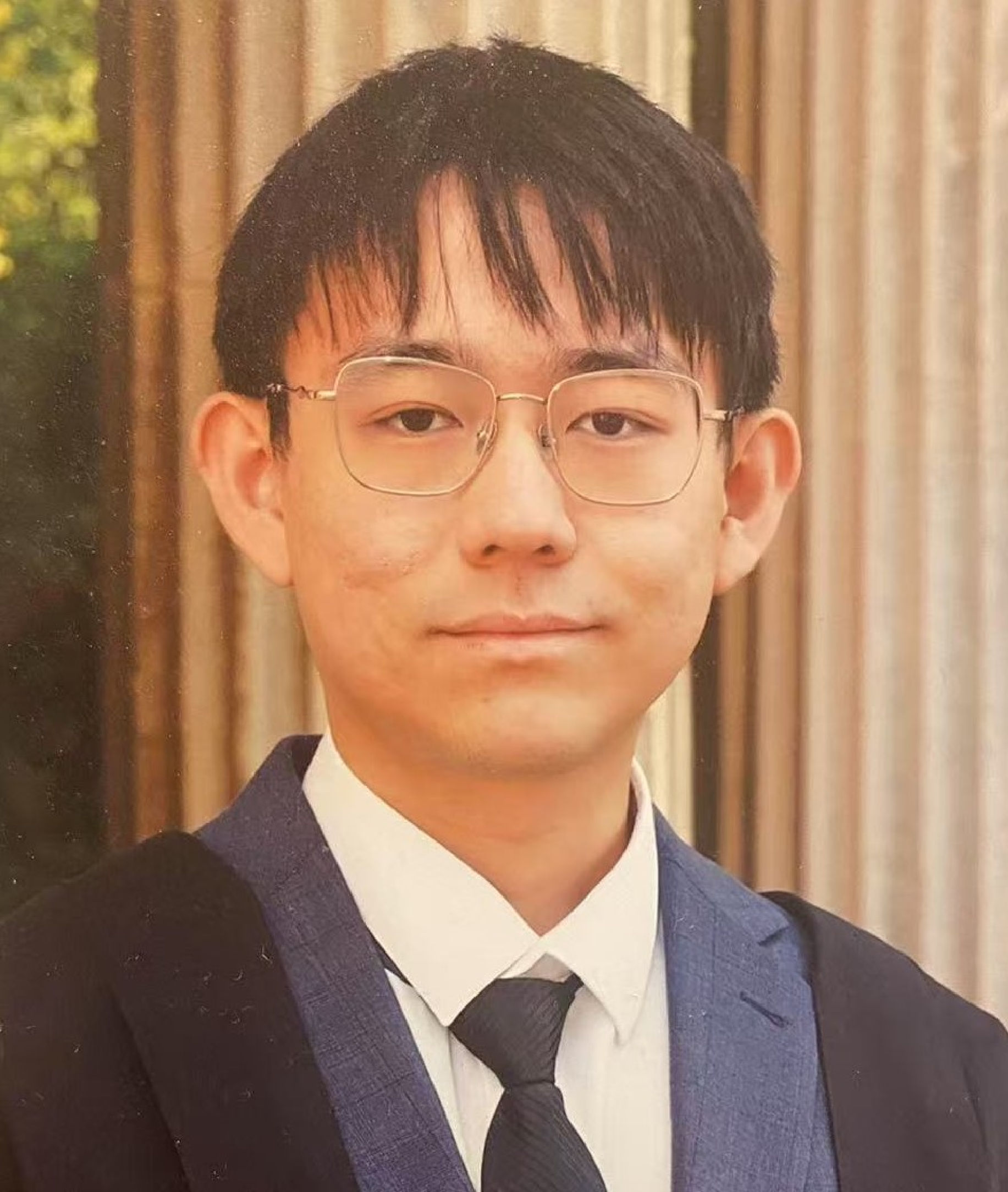}}]{Tiankuo Jiao}
received the B.Eng. degree in Electrical and Electronic Engineering with First Class Honors from the University of Liverpool, U.K., in 2023. He subsequently earned an M.Res. degree in Photonic and Electronic Systems from the University of Cambridge, U.K., in 2024. He is currently pursuing a Ph.D. at the LiFi Research and Development Center (LRDC), University of Cambridge, where his research focuses on the design and implementation of coherent optical wireless communication (OWC) systems and the development of advanced digital signal processing (DSP) algorithms for next-generation light fidelity (LiFi) connectivity.
\end{IEEEbiography}


\begin{IEEEbiography}[{\includegraphics[width=1in,height=1.25in,clip,keepaspectratio]{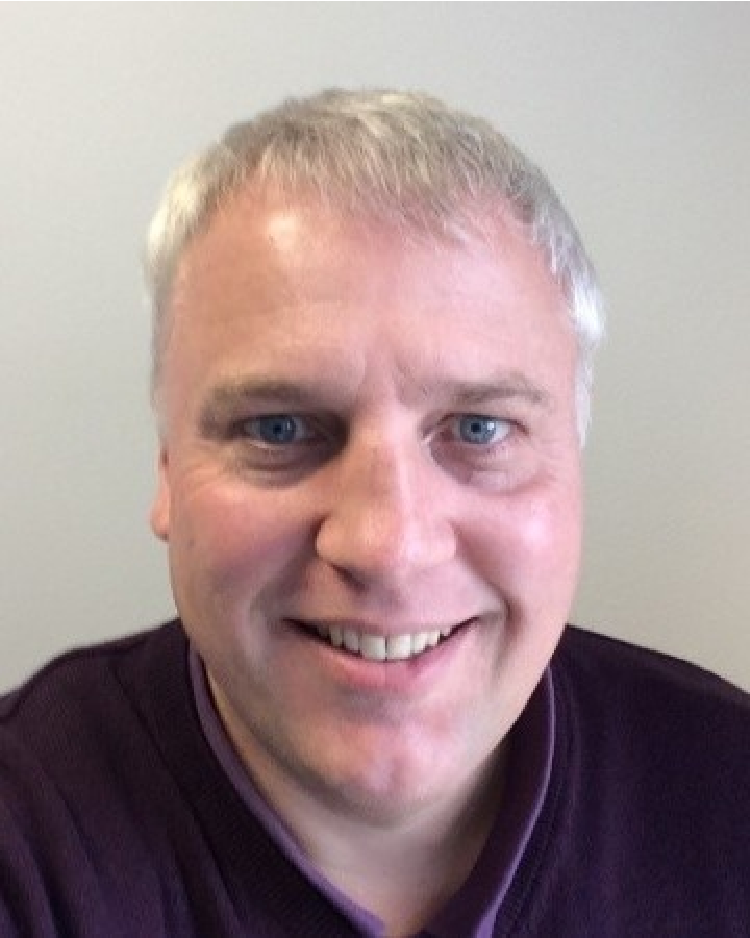}}]{David Butler} 
received the M.Eng. degree in Electrical and Electronic Engineering from the University of Bath, U.K., in 1991. He is currently a Senior R\&D Engineer at BBC Research and Development, where he has over 25 years of experience in the communications and broadcasting industry. His work encompasses the design, prototyping, and deployment of end-to-end systems, with particular emphasis on analog and radio frequency (RF) technologies. He has contributed extensively to the development of broadcast infrastructure and has played a key role in advancing technical innovations within the BBC. His areas of specialization include RF systems, signal processing, and system integration across the full product life cycle.
\end{IEEEbiography}


\begin{IEEEbiography}[{\includegraphics[width=1in,height=1.25in,clip,keepaspectratio]{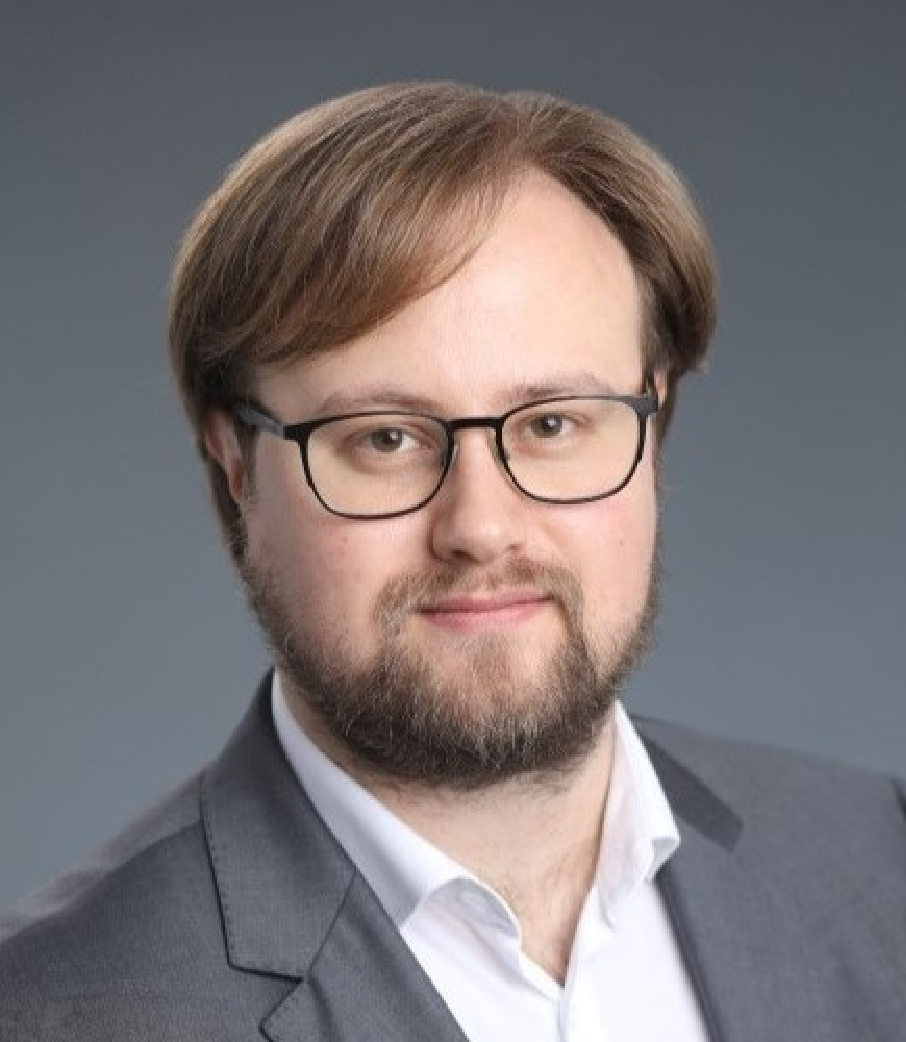}}]{Nikolay Ledentsov Jr.} 
received his B.Sc. and M.Sc. degrees from the Technical University of Berlin, Germany, in 2012 and 2014, respectively, where his research focused on the growth and characterization of Indium Aluminum Gallium Nitride (InAlGaN) green and ultraviolet (UV) light emitting diodes (LEDs). He completed his Ph.D. at the Technical University of Warsaw, Poland, in 2023, focusing on high-speed data transmission with intrared (IR) vertical-cavity surface-emitting lasers (VCSELs). He was a Senior Engineer at VI Systems GmbH, where he was responsible for the research and development of VCSELs for high-speed optical links, and manufacturing and characterization of light emitters and photodiodes. He is currently head of the characterization and VCSEL development at EPIGAP OSA Photonics GmbH.
\end{IEEEbiography}


\begin{IEEEbiography}[{\includegraphics[width=1in,height=1.25in,clip,keepaspectratio]{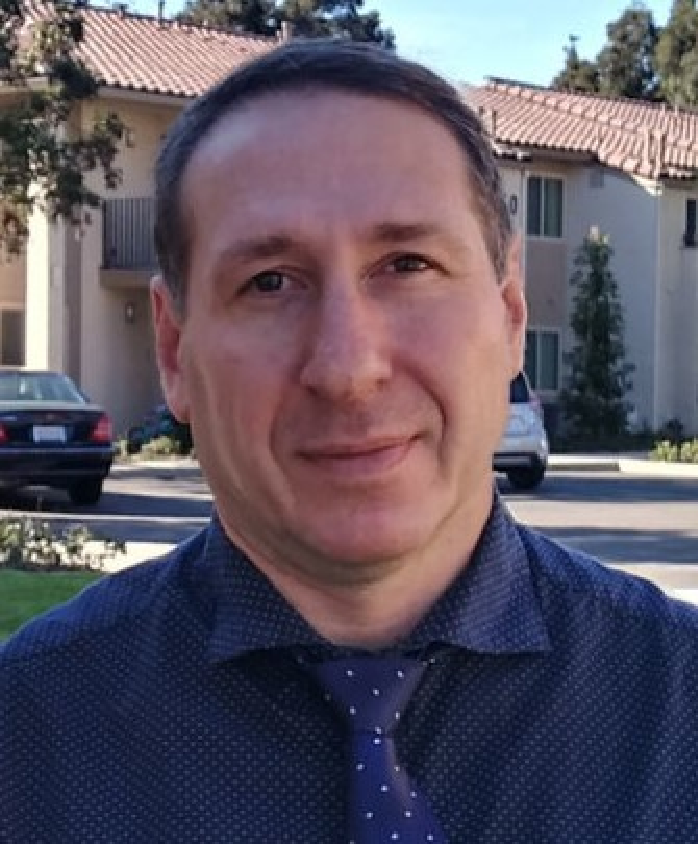}}]{Ilya Titkov} 
received his Ph.D. degree in Semiconductor Manufacturing Technology from Peter the Great St. Petersburg Polytechnic University, Saint Petersburg, Russia, in 2000. From 2002 to 2009, he worked as a researcher at the Ioffe Institute’s Laboratory of Semiconductor Devices Physics in Saint Petersburg, Russia, where he focused on semiconductor material physics and oxide-based device structures. Since 2014, he has held a research position at the Aston Institute of Photonic Technologies, Aston University, Birmingham, U.K., contributing to LED efficiency analysis and photonic device development. He currently serves as R\&D Manager and Senior Researcher at VI Systems GmbH, specializing in advanced characterization methods and failure analysis for high-speed VCSEL technologies and optoelectronic devices.
\end{IEEEbiography}


\begin{IEEEbiography}[{\includegraphics[width=1in,height=1.25in,clip,keepaspectratio]{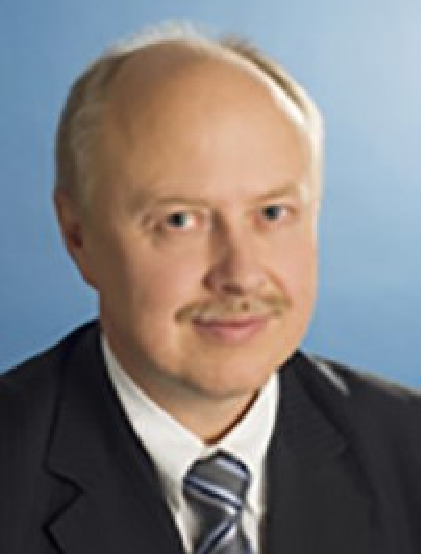}}]{Nikolay Ledentsov} 
(Senior Member, IEEE) received the Diploma in electrical engineer from the Electrical Engineering Institute, Leningrad, Russia, in 1982, the Ph.D. ( Cand. Sci.) and Dr. Sci. (Habil.) degrees in physics and mathematics from A. F. Ioffe Institute, Leningrad/Saint Petersburg, Russia, in 1987 and 1994, respectively. Since 1994, he has been a Professor of electrical engineering and, since 2005, a certified Professor of physics and mathematics with Ioffe Institute. During 1996--2007, he was a Professor with the Technical University of Berlin, Berlin, Germany. He has authored or coauthored more than 900 papers in technical journals and conference proceedings and 38 patent families. His Hirsch factor is 83. His current research interests include physics and technology of semiconductor nanostructures and design and technology of advanced optoelectronic devices. He is a Fellow of the Institute of Physics and a Member of the Russian Academy of Sciences. He was the recipient of the Young Scientist Award from the International Symposium on Compound Semiconductors in 1996 for outstanding contributions to the development of physics and MBE growth of InGaAs/GaAs quantum dots structures and quantum dot lasers, State Prize of Russia for Science and Technology in 2001, Prize of the Berlin Brandenburg Academy of Sciences in 2002, and other awards and recognitions. During 1995--1996, he was the recipient of Alexander von Humboldt Fellowship. Since 2006, he has been the Chief Executive Officer of VI Systems GmbH.
\end{IEEEbiography}


\begin{IEEEbiography}[{\includegraphics[width=1in,height=1.25in,clip,keepaspectratio]{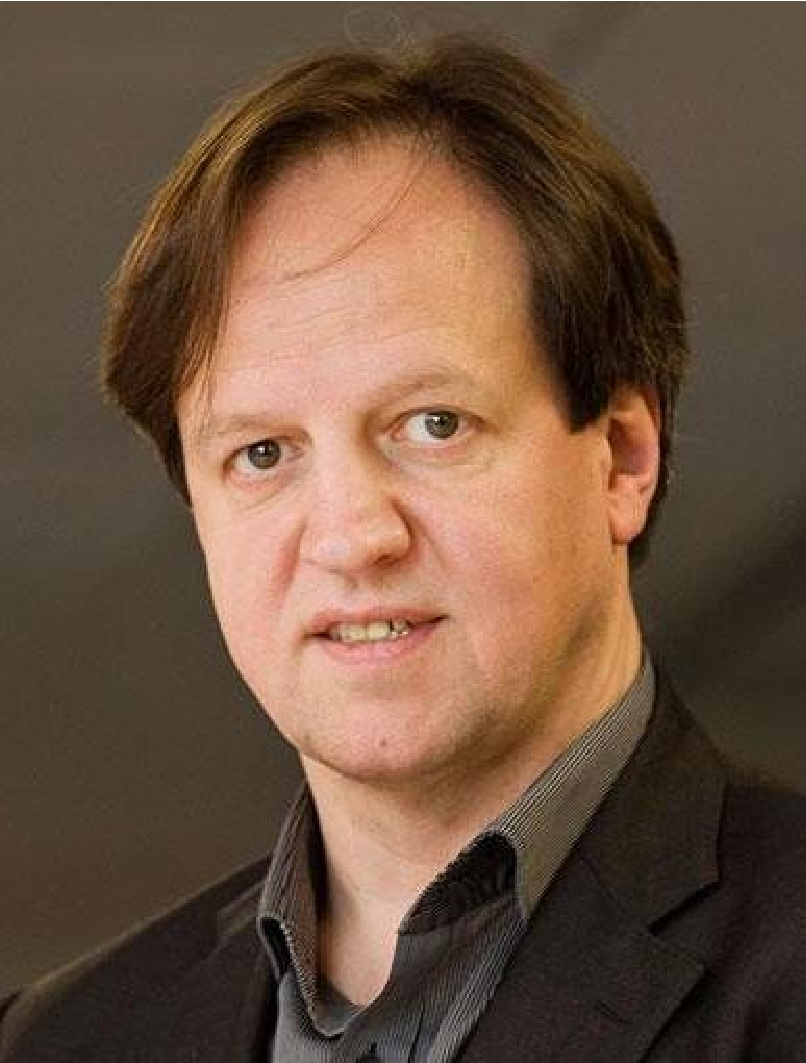}}]{Harald Haas} 
(Fellow, IEEE) received his Ph.D. from The University of Edinburgh, U.K., in 2001. He is the Van Eck Chair of Engineering at the University of Cambridge and the founder of pureLiFi Ltd., where he also serves as the Chief Scientific Officer (CSO). His recent research interests focus on photonics, communication theory and signal processing for optical wireless communication systems. Since 2017, he has been recognised as a highly cited researcher by Clarivate/Web of Science. He has delivered two TED talks and one TEDx talk. In 2016, he received the Outstanding Achievement Award from the International Solid State Lighting Alliance. He was awarded the Royal Society Wolfson Research Merit Award in 2017, the IEEE Vehicular Technology Society James Evans Avant Garde Award in 2019, and the Enginuity: The Connect Places Innovation Award in 2021. In 2022, he received the Humboldt Research Award for his research contributions. He is a Fellow of the Royal Academy of Engineering (RAEng), the Royal Society of Edinburgh (RSE), and the Institution of Engineering and Technology (IET). In 2023, he was shortlisted for the European Inventor Award.
\end{IEEEbiography}

\end{document}